\documentstyle[11pt,aaspp4,psfig,epsf]{article}
\begin{document}

\title{Deconvolution with correct sampling}

\author{P.~Magain\altaffilmark{1}}
\affil{\small  Institut d'Astrophysique,  Universit\'{e}  de
Li\`{e}ge,\\ 5, avenue de Cointe,\\ B--4000 Li\`ege Belgium\\
email: magain@astro.ulg.ac.be}

\author{F.~Courbin}
\affil{\small  Institut d'Astrophysique,  Universit\'{e}  de
Li\`{e}ge,\\ 5, avenue de Cointe,\\ B--4000 Li\`ege Belgium\\
email: courbin@astro.ulg.ac.be\\
\vspace*{3mm}
\small  URA  173  CNRS-DAEC, Observatoire de Paris,\\
F-92195 Meudon Principal CEDEX, France}

\author{S.~Sohy}
\affil{\small  Institut d'Astrophysique,  Universit\'{e}  de
Li\`{e}ge,\\ 5, avenue de Cointe,\\ B--4000 Li\`ege Belgium\\
email: sohy@astro.ulg.ac.be}

\altaffiltext{1}{Ma\^{\i}tre  de Recherches  au  Fonds  National de la
Recherche Scientifique (Belgium)}

\begin{abstract}

A new method  for improving the resolution  of astronomical images  is
presented.  It  is based on the  principle that sampled data cannot be
fully deconvolved without  violating the sampling theorem.   Thus, the
sampled image  should  not be deconvolved  by  the  total Point Spread
Function, but by a narrower function  chosen so that the resolution of
the deconvolved image is compatible with the adopted sampling.

Our deconvolution method gives  results which are, in at least some cases,
superior to those of  other commonly used techniques$\,$: 
in  particular, it does not
produce  ringing   around  point  sources  superimposed on    a smooth
background.  Moreover, it  allows  to perform accurate  astrometry and
photometry of crowded fields.  These improvements are a consequence of
both  the correct treatment of sampling   and the recognition that the
most probable astronomical image is not a flat one.

The  method  is   also well   adapted to  the  optimal combination  of
different images  of the same object,   as can be obtained,  e.g., from
infrared observations or via adaptive optics techniques.

\end{abstract}

\pagebreak

\section{Deconvolution}

Any recorded  image is blurred  whenever the instrument used to obtain
it has a finite resolving power$\,$: for example, the image of a point
source seen through a telescope has an angular size which is inversely
proportional to the diameter of the primary mirror.  If the instrument
is ground-based, the image is  additionally degraded by the  turbulent
motions in the earth's atmosphere.

Much effort is  presently devoted  to  the improvement  of the spatial
resolution of astronomical images, either via  the introduction of new
observing techniques  (e.g.\  interferometry or  adaptive optics,  see
L\'ena  1996,  Enard  et al.\  1996)  or via    a subsequent numerical
processing  of the image (deconvolution).   It  is, in fact, of  major
interest to combine both methods to reach an even better resolution.

An observed    image  may usually be  mathematically   expressed  as a
convolution   of the original     light distribution with the  ``total
instrumental  profile'' --- the latter   being  the image  of a  point
source obtained   with  the   instrument  considered,  including   the
atmospheric  perturbation   ({\em  seeing})   if relevant.  The  total
blurring function is called  the {\em Point  Spread Function} (PSF) of
the image.

Thus, the imaging equation may be written$\,$:
\begin{equation}
d(\vec{x}) = t(\vec{x}) \ast f(\vec{x}) + n(\vec{x})
\end{equation}
where   $f(\vec{x})$ and $d(\vec{x})$  are  the  original and observed
light distributions,  $t(\vec{x})$  is the total PSF  and $n(\vec{x})$
the measurement errors ({\em noise}) affecting the data.

In addition, on  all modern light  detectors (e.g.\  CCDs whose pixels
have finite dimensions),  the  observed  light  distribution is   {\em
sampled},  i.e.\ known only at regularly  spaced sampling points.  The
imaging equation for a sampled light distribution then becomes$\,$:
\begin{equation}
d_i = \sum_{j=1}^{N} t_{ij} \, f_j + n_i
\end{equation}
where $N$  is the number  of sampling points,  $d_j$, $f_j$, $n_j$ are
vector  components   giving   the  sampled   values  of  $d(\vec{x})$,
$f(\vec{x})$,  $n(\vec{x})$ and $t_{ij}$ is the  value at point $j$ of
the PSF centered on point $i$.

The aim of deconvolution may be stated in the following way$\,$: given
the observed image $d(\vec{x})$  and the PSF $t(\vec{x})$, recover the
original light distribution $f(\vec{x})$.   Being an  inverse problem,
deconvolution is also an ill-posed problem, and no unique solution can
be found,  especially in the presence  of noise.  This  is due  to the
fact that  many light  distributions  are, after convolution  with the
PSF,  compatible within  the  error  bars  with  the  observed  image.
Therefore, regularization techniques  have  to  be  used  in order  to
select a plausible solution amongst the  family of possible ones and a
large  variety of deconvolution  methods have been proposed, depending
on the way this particular solution is chosen.

A typical method  is to   minimize the  $\chi^2$  of the   differences
between  the  data and    the   convolved model, with   an  additional
constraint imposing smoothness   of the solution.   $f_j$ is  then the
light distribution which minimizes the function$\,$:
\begin{equation}
{\cal S}_1 = \sum_{i=1}^{N} \frac{1}{\sigma_i^2} \left [ \sum_{j=1}^{N}
t_{ij} f_j - d_i \right ]^2 \; + \; \lambda \, H(f_1,\cdots,f_N)
\end{equation}
where the  first term in the sum  is the  $\chi^2$ with $\sigma_i$ the
standard deviation of the image intensity measured at the $i^{\rm th}$
sampling point, $H$ is a  smoothing function and $\lambda$ a  Lagrange
parameter  which   is determined so  that   the reconstructed model is
statistically  compatible  with the  data   ($\chi^2  \simeq N$).   If
$H(f_1,\cdots,f_N) =  \sum_{i=1}^{N} p_i   \ln p_i$, with    $p_i$ the
normalized flux at  point $i$, $p_i =  f_i / \sum_{j=1}^{N}  f_j$, one
obtains the   so-called  {\em   maximum  entropy  method} for    image
deconvolution (Narayan and Nityananda 1986, Skilling and Bryan 1984).

In  order  to choose   the correct answer   in  the family of possible
solutions to this inverse problem, it  is also very useful to consider
any available {\em prior knowledge}.  One such  prior knowledge is the
{\em positivity} of the light distribution$\,$: no negative light flux
can be recorded,  so that all  solutions with  negative values  may be
rejected.  The maximum entropy method automatically ensures positivity
of the solution.  This is also the case, under certain conditions, for
other popular methods, such as the Richardson-Lucy iterative algorithm
(Richardson 1972, Lucy 1974).

Most of  the known deconvolution  algorithms  suffer from  a number of
weak  points which strongly    limit their usefulness.   The  two most
important   problems in   this respect    are  the following$\,$:  (1)
traditional deconvolution  methods tend to  produce artefacts  in some
instances  (e.g.\   oscillations    in    the  vicinity     of   image
discontinuities,   or around point  sources  superimposed  on a smooth
background)$\,$; (2) the  relative  intensities of different  parts of
the  image (e.g.\ different stars) are  not conserved, thus precluding
any   photometric measurements.  In the   next sections, we identify a
plausible cause of these problems and show how to circumvent it.

\section{Sampling}

The sampling theorem (Shannon 1949, Press et al.\ 1989) determines the
maximal sampling interval  allowed so that  an entire function  can be
reconstructed  from sampled  data.  It states   that a  function whose
Fourier  transform  is  zero  at  frequencies   larger  than  a cutoff
frequency  $\nu_0$  is fully   specified  by  values  spaced at  equal
intervals not exceeding $(2\,\nu_0)^{-1}$.  In practice, for functions
whose Fourier   transform does not   present such a  cutoff frequency,
$\nu_0$ may  be taken as the   highest frequency at  which the Fourier
transform emerges from the noise.

The  imaging instruments are  generally  designed so that the sampling
theorem is approximately fulfilled in average observing conditions.  A
typical sampling encountered is  $\sim 2$ sampling intervals  per Full
Width at  Half  Maximum (FWHM) of   the PSF  (this does certainly  not
ensure good sampling   for high signal-to-noise  (S/N)  images, but is
roughly sufficient at low S/N).

The main  problem   with  classical deconvolution   algorithms  is the
following$\,$: {\em if the observed data are sampled so that they just
obey the sampling theorem, the deconvolved data will generally violate
that same     theorem}.  Indeed,   increasing  the  resolution   means
recovering  highest Fourier frequencies,  thus  increasing the  cutoff
frequency, so that the correct sampling becomes denser.

One might  object  that some deconvolution  algorithms, which  allow a
different sampling in    the deconvolved image,   could overcome  this
problem$\,$:  it  would be possible   to  keep a  correct  sampling by
shortening  the sampling  interval.   This   is however an    illusory
solution, since the only limit on the  frequency components present in
an arbitrary image comes from the PSF of the instrument used to record
it.   Removing the  effect  of the PSF  would   allow the presence  of
arbitrary high  frequency components,  and  thus an  infinitely  small
sampling interval would have to be used.

This is particularly  true if the  image contains point sources, which
is generally the case  for  astronomical images.  Indeed,  the angular
diameters of  most stars  ($\ll 0.001\,  {\rm  arcsec}$) are  so small
compared  to  the sampling interval  ($\sim  0.1\, {\rm arcsec}$) that
they may be considered as point  sources (``$\delta$-functions'').  In
such an instance, it would be hopeless to reduce the sampling interval
in an attempt to obtain a good sampling of such ``$\delta$-functions".

This is  one of the  sources of some of the  artefacts  present in the
deconvolved images and, in particular, of the ``ringing'' around point
sources superimposed  on a diffuse   background.   The origin of  this
``ringing'' may be understood in the following way.

Let us assume that we have a continuous (i.e.\ not sampled) noise-free
image  of a  field  containing point  sources,  and  observed with  an
instrument having  a  known PSF.   For  simplicity,  we restrict   our
considerations to   one-dimensional   images.  If   we  can  perfectly
deconvolve this image, we shall obtain a solution $f(x)$ in which each
point source is represented by a Dirac $\delta$-function.

Now,  let us assume  that we have the same   image, but sampled on $N$
points, with a sampling step $\Delta x$.  The Fourier transform of its
deconvolution  may be obtained from the  Fourier transform $F(\nu)$ of
the  continuous deconvolution  in     the following way$\,$:    repeat
periodically $F(\nu)$ with the Nyquist frequency $\nu_{Ny} = (2 \Delta
x)^{-1}$,  take  the  sum of all   these  periodical replicas  at each
frequency point,  isolate one period  and  sample  it on  $N$  equally
spaced frequency points.

Isolating   one period means  multiplying the  Fourier  transform by a
rectangular (``box'') function which equals 1 in an interval of length
$\nu_{Ny}$  and 0 outside.  Now,  a   convolution in the image  domain
translates  into   a  simple product  in     the Fourier domain,   and
vice-versa.  This multiplication   by  a box function  in  the Fourier
domain is thus  equivalent, in the image domain,  to  a convolution by
the Fourier transform of the box function, which  is a function of the
form $\sin x /  x$.  The solution of  the deconvolution problem for  a
sampled image with  point sources is thus  the (sampled) convolution of
the exact solution  $f(x)$ with a function of  the form $\sin x  / x$.
Each $\delta$-function is thus  replaced by an oscillatory $\sin
x / x$ function, which explains the ringing.

Another,   more intuitive, explanation  of   the  same effect is   the
following.  If a point  source is located  between two sampling points
(as will generally be  the case), in  order to correctly reproduce its
position, the deconvolution   algorithm  will have to   distribute its
intensity  over several sampling points.  But,  then, the width of the
source will be  too  large and  ringing will  appear as the  algorithm
attempts to decrease  the intensity on the  edges of the reconstructed
source, in order to  keep the convolved model  as close as possible to
the observed data.

In fact, it  is not possible to correctly  reproduce both the position
and the width of a sampled point source.  To reproduce the zero width,
the full signal must be  concentrated on a single  sampling point.  On
the other hand, to  reproduce the position  with a precision which  is
better  than the sampling interval, the  signal  has to be distributed
over several points.

\section{Solution}

The correct  approach  to this sampling problem  is  thus {\em not} to
deconvolve with the total PSF $t(\vec{x})$, but rather with a narrower
function $s(\vec{x})$ chosen so that the deconvolved image has its own
PSF $r(\vec{x})$ compatible  with  the adopted sampling.  These  three
functions are simply related by$\,$:
\begin{equation}
t(\vec{x}) = r(\vec{x}) \ast s(\vec{x})
\end{equation}

Note that a similar decomposition was proposed, in a completely
different context (reduction of artefacts in maximum-likelihood
reconstructions for emission tomography), by Snyder et al.\  (1987).

The  shape and width  of $r(\vec{x})$ can be chosen  by the user.  The
only  constraint is that Eq.~(4)  admits a solution $s(\vec{x})$.  The
function  $s(\vec{x})$ by which     the  observed  image has     to be
deconvolved is thus  obtained as the   deconvolution of the total  PSF
$t(\vec{x})$  by the final PSF  $r(\vec{x})$.  Of course, the sampling
interval of the deconvolved image  does not need  to  be equal to  the
sampling interval of the original  image, so that $r(\vec{x})$ may  be
much narrower than  $t(\vec{x})$, even if  the original sampling would
not allow it.  Choosing a sufficiently narrow $r(\vec{x})$ effectively
insures that Eq.~(4) will admit a solution $s(\vec{x})$.  Note also that,
contrary to other traditional methods (the success of which depends
crucially on the effectiveness of the positivity constraint), we have
no positivity constraint on the PSFs $r(\vec{x})$ and $s(\vec{x})$.

Thus, the deconvolution algorithm should not  attempt to determine the
light   distribution as if    it were  obtained with   an  {\em ideal}
instrument (e.g.\ a space telescope  with a primary mirror of infinite
size).  This is forbidden  as long as the  data are sampled.   Rather,
the aim of deconvolution should be to determine the light distribution
as if it were observed with a {\em better} instrument (e.g.\ a 10$\,$m
space telescope).

Deconvolution    by $s(\vec{x})$ ensures  that  the  solution will not
violate the sampling theorem.  It also has a very important additional
advantage$\,$: if the image contains point sources, their shape in the
deconvolved image is   now     precisely known$\,$:  it     is  simply
$r(\vec{x})$.  This is a very strong {\em prior knowledge}, and it may
be used to constrain  the  solution  $f(\vec{x})$,  which can  now  be
written$\,$:
\begin{equation}
f(\vec{x}) = h(\vec{x}) + \sum_{k=1}^{M} a_k \; r(\vec{x}-\vec{c}_k)
\end{equation}
where $M$   is  the number of  point   sources,  for which   $a_k$ and
$\vec{c}_k$ are free parameters corresponding to their intensities and
positions, and $h(\vec{x})$ is the extended component of the solution,
i.e.\ generally a rather smooth background.

We  can use another prior knowledge  to constrain the solution$\,$: we
know  that  the background $h(\vec{x})$ can   also  be written  as the
convolution  of  some  function $h'(\vec{x})$   with   the PSF of  the
solution $r(\vec{x})$ $\,$:
\begin{equation}
h(\vec{x}) = r(\vec{x}) \ast h'(\vec{x})
\end{equation}

However, we cannot   use  that decomposition directly  and   determine
$h'(\vec{x})$ instead  of  $h(\vec{x})$ because   $h'(\vec{x})$  might
violate the sampling   theorem, even  if  it  does  not contain  point
sources.  Rather, we  may use this  knowledge to impose smoothness  of
$h(\vec{x})$ on the scale length of $r(\vec{x})$.

So, instead of regularizing the solution by a  global function such as
the entropy,   we   use a  function    imposing  local smoothness   of
$h(\vec{x})$ on  the known scale  length.  We thus choose the solution
which minimizes the function$\,$:
\begin{equation}
{\cal S}_2 = \sum_{i=1}^{N} \frac{1}{\sigma_i^2} \left [ \sum_{j=1}^{N}
s_{ij} \left ( h_j + \sum_{k=1}^{M} a_k \, r(\vec{x}_j \! - \! \vec{c}_k)
\right )
- d_i \right ]^2 \; + \; \lambda \, \sum_{i=1}^{N} \left [ h_i -
\sum_{j=1}^{N} r_{ij} \, h_j \right ]^2
\end{equation}
with respect to the unknowns $h_i \, (i = 1,\ldots,N)$, $a_k$ and
$\vec{c}_k \, (k = 1,\ldots,M)$.

Although the smoothing term in the right-hand  side  of Eq.~(7) will
not force $h(\vec{x})$ to exactly obey Eq.~(6), it essentially
contains the Fourier components  with frequencies higher than those of
the deconvolved PSF $r(\vec{x})$.  Minimizing this term will force the
background component to  contain only the frequencies compatible  with
$r(\vec{x})$ and, thus, with the adopted sampling.

One additional   improvement  may be    introduced.  In  general,  the
Lagrange multiplier  $\lambda$  is chosen  so  that $\chi^2 \simeq N$.
This ensures that the fit is statistically correct globally.  However,
some  regions  of the  image   may be overfitted,   and  others may be
underfitted.    In practice, this will    generally  be the  case$\,$:
although the residuals   will  be correct  on the  average,  they will
systematically be too small  in some parts of  the image and too large
in other parts.

To avoid this problem, one may replace the smoothing function by$\,$:
\begin{equation}
H(f_1,\cdots,f_N) = \sum_{i=1}^{N} \lambda_i \left [ h_i -
\sum_{j=1}^{N} r_{ij} \, h_j \right ]^2
\end{equation}
where $\lambda_i$ is the value at the $i^{\rm th}$ sampling point of a
function $\lambda(\vec{x})$  which is chosen so  that the residuals of
the  fit are correct   locally,   i.e.\ that they  are   statistically
distributed with the correct standard deviation in any sub-part of the
image.

% New 
In practice, an image of the square of the normalized residuals (observed
data minus convolved model, divided by sigma) is computed and then 
smoothed with an appropriate function, so that any value is replaced
by a weighted mean on a neighborhood containing a few dozens of pixels.
The parameter $\lambda$ is then adjusted until this image is close to
one everywhere.

\section{Examples}

Our deconvolution    program  implements the    ideas exposed  in  the
preceding section.  The light distribution aimed  at is written as the
sum of a  smooth  background plus  a number   of point sources.    The
sampling step of the deconvolved image is chosen, as well as the final
PSF $r(\vec{x})$, compatible with this sampling  (in general, we adopt
a gaussian function, with  a few pixels  FWHM).  Approximate values of
the unknowns   are chosen and the  function  ${\cal S}_2$ is computed,
together with its   derivatives with respect to   all  variables.  The
minimum  of  ${\cal S}_2$ is  then  searched for,   using an algorithm
derived from  the classical conjugate  gradient method (Press  et al.\
1989).  The fit's  residuals are  then computed  and a check  of their
statistical correctness is performed.   If this test is not satisfied,
the  Lagrange multiplier    $\lambda$   is  replaced by  a    variable
$\lambda(\vec{x})$ which is varied until the  residuals conform to the
statistical expectations.

% New
The present version of the program runs on PCs and workstations and can
handle images of reasonable size (e.g., 256 $\times$ 256 pixels)
containing up to several hundreds of stars.  The main weakness of the present
implementation is related to the conjugate gradient algorithm, which
is not always able to find the global minimum of the function, especially
when the number of point sources is large.  We are presently working on a 
new optimization technique which would allow our method to be applied
to the photometry of crowded fields with thousands of stars.

It may not seem obvious at first sight to select the correct number of
point sources to be included in the  solution.  However, the algorithm
allows to constrain  this number in  a very efficient  way$\,$: if too
few  point sources  are  entered, it  will generally be  impossible to
obtain statistically correct  residuals  locally, in all  sub-parts of
the image.  On   the  other  hand, if  too   many point  sources   are
considered,  the algorithm will  either attribute essentially the same
position  or   negligible  intensities  to    several  of  them.   Our
methodology is thus to model the data with the minimum number of point
sources necessary to yield statistically  correct residuals locally in
all sub-parts of the image, in the sense described at the end of the
preceding section.

Figure  1 compares the results of   our new deconvolution algorithm to
those of  three  classical methods in  the  case of  a  simulated star
cluster partly  superimposed on a  smooth background  (e.g.\ a distant
elliptical galaxy).  The input point sources were selected from the observed
image alone, without any prior knowledge of the exact solution.  It is
clear that our result is free from the artefacts  present in the other
methods and that it allows  an accurate reconstruction of the original
light  distribution.  Another important  property  of our technique is
that it allows, contrary to the other ones, an accurate measurement of
the  positions and intensities of  the point sources.  This point will
be discussed more extensively in the next section.

An application   to real astronomical  data is  shown on Fig.~2, which
displays     a mediocre resolution  image     of the ``Cloverleaf'', a
gravitationally lensed quasar (Magain et al.\ 1988), together with the
deconvolved version, using  a sampling interval  twice as short.   The
four lensed  images, which were unresolved  in the  original data, are
completely  separated after  deconvolution.   The  deduced fluxes  are
fully compatible with those measured  on higher resolution images and,
although the original resolution is 1.3 arcsec only and the pixel size
is 0.35  arcsec, the  deduced  image  positions are accurate  to  0.01
arcsec.

Figure 3 illustrates the deconvolution of an image of the compact star
cluster Sk$\,$157 in the  Small Magellanic Cloud (Heydari-Malayeri  et
al.\  1989).   The original image was   obtained with the ESO/MPI 2.2m
telescope at La Silla, in average seeing conditions (1.1 arcsec FWHM).
While the original  maximum entropy deconvolution (Heydari-Malayeri et
al.\ 1989) allowed to resolve the cluster into  12 components, our new
algorithm detects, from the same input data, more than 40 stars in the
corresponding area.

Another important   application of our   algorithm is the simultaneous
deconvolution of different images of the same field.  These images may
be  obtained  with the same instrument  or  with different  ones.  The
solution is then a light distribution which is compatible with all the
images considered.   Our   technique even allows  to   let,  e.g., the
intensities  of the point sources  converge to different values in the
different images,  so that variable objects may  be  considered.  This
technique  should be very   useful for the  photometric monitoring  of
variable  objects   in crowded fields   or  overimposed  on  a diffuse
background (e.g.\ Cepheids in distant galaxies, gravitationally lensed
QSOs,...).

Figure 4   illustrates this  simultaneous  deconvolution  on simulated
images,  the first of which  has a good resolution but  a poor S/N (as
might be obtained  with a space telescope)   and the second one a  low
resolution and a high  S/N (a typical  image from a large ground-based
telescope).  Contrary to Lucy's method  (Hook and Lucy 1992) which is 
very  sensitive
to  the noise present in  one of the  images,  our technique allows to
reliably recover both the  high resolution of the  space image and the
hidden information content of the ground-based one.

In the same spirit, our algorithm is well adapted to the processing of
images obtained  with infrared or  adaptive optics  techniques.   
In the  latter,
numerous  short exposures of the same  field are usually obtained, the
shape of the  mirror    being  continuously adapted to   correct   for
atmospheric distortions.  So, the  observations consist in a number of
images of the same field, each of them having its own PSF.  Performing
a  simple sum results in an  image whose spatial resolution is typical
of    the  average   observing   conditions,  while    a  simultaneous
deconvolution not  only allows to take  count  of the best conditions,
but even results in an improved resolution  by optimally combining the
information content of the different images.  A simple illustration of
these considerations  is   provided  by    Fig.~5,  which shows    the
simultaneous deconvolution of  four adaptive-optics-like images of the
same field, where the PSF as well as the image centering vary from one
observation to the other.  Of course, the PSF needs to be known for
each individual observation, but only with an accuracy comparable to
that of the observation itself.

\section{Astrometric and photometric accuracy}

Traditional deconvolution  methods   are notoriously  unable  to  give
photometrically   accurate results.  Two    main reasons for that  are
readily identified.

First, as  we have already mentioned,  these methods generally produce
rings when point sources are overimposed on  a diffuse background.  In
fact, these rings tend to appear  as soon as the positivity constraint
is inefficient  to  inhibit them,  that  is, as soon   as some flux is
distributed around the point sources.  This is  most clearly seen when
this flux  is in the form  of a smooth  background, but  the effect is
also present  if  the flux is  distributed  among, e.g.,  a number  of
fainter  stars.  In this  case, the  rings  around the star considered
will interfere with the intensity in the neighbouring sources, and the
photometry of the latter ones will be affected.

A second photometric  bias comes from the fact  that, among the family
of   possible solutions   to  the  inverse   problem,  most  classical
algorithms select, in one way or another,  the smoothest one according
to some criterion.   These  algorithms thus  produce images where  the
peaks  corresponding  to point sources  deviate  as little as possible
with respect to the background --- provided, of course, that the model
fits   the data.   This   implies a  systematic  underestimate of  the
intensity peaks, and thus, a photometric bias.

An example of  these effects is illustrated by  the  deconvolution of an
image  of  two point  sources with  varying separation.   A simple
image  was constructed,  with  two point sources  having an  intensity
ratio of  0.1, and convolved  with a  Gaussian PSF  of  7 pixels FWHM,
plus some gaussian noise so that the peak S/N ratio reaches 100.
Figure  6  shows the deduced intensity ratio as a function of the source
separation, as  derived after deconvolution  with the maximum entropy
method.  It   clearly   shows that  the
photometry is not preserved, even when  the two stars are separated by
nearly two FWHMs.  For more details on the photometric accuracy
of deconvolution algorithms (in the special case of HST images), see
Busko (1994).

Our algorithm naturally avoids  these  two biases.  Indeed, the   fact
that the sampling theorem is obeyed in the deconvolved image, combined
with the  fact that no smoothing   of the point  sources is attempted,
naturally ensure that no ringing is present around the star peaks, and
that  no bias  will appear as   a  consequence of  smoothing.  This is
illustrated in Fig.~7,  which shows the  results of a photometric test
applied to a synthetic field containing 200 stars in a 128$\times$ 128
pixels image.  The positions and central  intensities were selected at
random,  and nearly all the stars  are blended to varying degrees (197
stars  out   of 200  have   the nearest   neighbour  within 2  FWHMs).
Moreover, these  stars  are superimposed  on   a variable  background.
Figure 7  clearly shows that no systematic  error is present, and that
the  intensities of  all but  the most  severely   blended objects are
reproduced with errors compatible with the photon noise.

Figure  8 illustrates the astrometric  accuracy of our algorithm.  For
the brightest stars, the positional  accuracy is generally better than
0.1 pixel, even in very severe blends.   The positions of stars with a
blend between 1 and 2 FWHM is generally  accurate within 0.02 pixel at
high S/N, and within 0.1 pixel otherwise.

There exits another deconvolution algorithm  which claims to achieve a
high photometric quality,  namely the so-called  {\em two-channel Lucy
method} (Lucy  1994, Hook  and  Lucy 1994).   As our  algorithm,   the
two-channel Lucy method is based on a decomposition of the deconvolved
model into point sources and background.

The main problem with that method is that, contrary to ours, the total
PSF is  used  in the deconvolution,  so  that the sampling problem  is
avoided only if each point source is exactly  centered on a pixel.  To
increase the accuracy, the  model can use  a finer pixel grid than the
data.   However, in high  S/N cases,  the  model pixels will generally
need to be very small if high accuracy is  aimed at (which is normally
the case  in high S/N observations...).  As  an example, let us recall
that the positions  derived from our  new algorithm for the  different
images of the  Cloverleaf gravitational lens  (Fig.~2) are accurate to
0.01 arcsec,   which is 1/35 of a   data pixel.   To achieve  the same
accuracy  with the  two  channel  Lucy  algorithm would  require  each
original pixel to be  devided in $\sim 35  \times 35 \sim 1000$  finer
pixels.    This  would rapidly  result      in huge data  frames   and
computationally intractable problems.

Another   weakness of the  two-channel Lucy  method  is that the point
source positions  have to  be  supplied  by the  user, and   cannot be
adjusted by the algorithm.  So,  no  astrometry can be performed  and,
moreover, in  the case of  high S/N data  with  many point sources, it
might require  an unreasonably large number of  trials for the user to
find a fairly good estimate of the source positions.

Finally, let us note that the user  has to choose arbitrarily not only
the number of iterations of the algorithm, but also a scale length for
the   smoothing of the     background  component.  These choices   are
generally made by looking at the results.  This approach can obviously
give  nice-looking   results, but  their  scientific soundness  may be
questioned.   On the contrary, the scale   length for the smoothing of
the background  in our method    is unambiguously  fixed by  the   PSF
$r(\vec{x})$ of the deconvolved image.

A comparison of the two-channel   Lucy  method with our algorithm   is
illustrated  below by  an example  which  is  not  meant  to provide a
general  comparison between  the  two methods,  but only to illustrate
some of the points in the preceding discussion.

Figure 9 shows the  deconvolution  of simulated data  containing three
point sources  superimposed on a  background which varies rather fast.
The peak S/N of 170 is quite reasonable  for modern CCD detectors.  In
order  to obtain  a satisfactory  result   with the two-channel   Lucy
method, each original  pixel was divided  into 16 model pixels and the
positions were adjusted iteratively   by the user.  In contrast,  when
running our  algorithm, we  kept the  same  pixel size as in  the data
(this is  why the results of  the two-channel method seem  smoother in
Fig.~9).  The deduced  background light distributions are compared  in
Fig.~10,  which   also shows   the difference   between these  deduced
backgrounds and the known solution, reconvolved to the same resolution
of 2 pixels FWHM.  It is immediately seen  that the residuals are much
less  important  in the case   of our  method (largest  residual~: 4.8
$\sigma$  as compared to  14  $\sigma$   with  the Lucy method,   mean
variance~:  1.8 $\sigma^2$ instead of  15 $\sigma^2$).  The photometry
of the point sources is also more accurate with  our method~: the mean
deviation is 1\%, as compared to 7\% with the two-channel method.

\section{Discussion}

We summarize here   some of the  reasons why   classical deconvolution
algorithms generally  give  rather disappointing results, and  why our
method allows to improve the situation.

A major  advantage of our method  over traditional ones comes from the
fact  that the deconvolved image  never violates the sampling theorem,
so  that the  fastest  image variations may  be correctly represented,
without the introduction of spurious rings, or Gibbs oscillations.

An additional drawback   of most traditional deconvolution  algorithms
lies in their smoothing  recipe.  For example,  in the maximum entropy
method, one assumes  that the most probable  image is a perfectly flat
one.  However, the most  probable astronomical image is {\em certainly
not} a flat  one.  It would rather look  like a dark background with a
number of   sharp sources.  Trying to    smooth the sharp   sources is
undesirable, and results in poor performance.

The  smoothing function used  in  the classical maximum entropy method
and most  of its  derivatives is,  moreover, a  {\em global} function,
i.e., a  function linking the  value of the  intensity in a particular
pixel to the values in all other pixels, even very remote ones.  Thus,
the flux distribution in one part of the image will  depend on what is
happening in  other remote parts (in  astronomical  images, this often
corresponds to  quite different parts of  the Universe).  This link is
obviously not based on physical grounds, and is totally avoided by our
smoothing function, which is purely local and linked to the PSF of the
deconvolved image.

Another weakness  of the most popular of  the classical methods (e.g.\
maximum  entropy or Richardson-Lucy) is  that the  solution depends on
the zero point level of the  image $\,$: this is  due to the fact that
the positivity constraint   is essential for   their success.  Indeed,
this positivity constraint is the main inhibitor of the ringing around
point sources:   by forbidding the   negative  lobes, it automatically
reduces the positive ones since the mean level must be compatible with
the observed data.  Adding a  constant to the  image data results in a
strong degradation of the performance  of these algorithms (which then
depend,   e.g., on a precise subtraction   of the sky  level).  On the
contrary,   our  technique is completely    independent of an additive
constant, and  it is reliable enough  that the  positivity constraint,
although it  can be used, is  not necessary in  most cases (it has not
been used in any of the examples shown in this paper).

As can be seen from  the above examples and  from the discussion,  our
new deconvolution technique   is well  adapted  to  the  processing of
astronomical images.  It  is however not restricted  to  that field of
imaging and, in fact, should be useful in several other areas where an
enhancement of  the image resolution  is desirable, or where different
images of the same object could be optimally combined.\\[3ex]

\parindent=0mm
\frenchspacing
\noindent
{\Large {\bf References}}\\

\normalsize

Busko, I.C., in {\em The Restoration of HST Images and Spectra-II},
R.J. Hanisch and R.L. White, eds., p.~279 (1994)

Enard, D., Mar\'echal, A. \& Espiard, J. {\em Rep. Prog. Phys.} {\bf 59},
601 (1996)

Heydari-Malayeri, M., Magain, P. \& Remy, M. {\em Astron. Astrophys.} {\bf 222},
41 (1989)

Hook, R.N. \& Lucy, L., {\em ST-ECF Newsletter} {\bf 17}, 10 (1992)

Hook, R.N. \& Lucy, L., in {\em The Restoration of HST Images and Spectra-II},
R.J. Hanisch and R.L. White, eds., p.~86 (1994)

L\'ena, P. {\em Astrophysique.  M\'ethodes physiques de l'observation},
2$^{\rm nd}$ edition, CNRS Editions, Paris, 1996 (english translation of the
first edition$\,$: {\em Observational Astrophysics}, Springer-Verlag, Berlin,
1988)

Lucy, L. {\em Astron. J.} {\bf 79}, 745 (1974)

Lucy, L. {\em ST-ECF Newsletter} {\bf 16}, 6 (1991)

Lucy, L., in {\em The Restoration of HST Images and Spectra-II}, R.J. Hanisch
and R.L. White, eds., p.~79 (1994)

Magain, P., Surdej, J., Swings, J.-P. {\em et al. Nature} {\bf 334},
325 (1988)

Narayan, R. \& Nityananda, R. {\em Annu. Rev. Astron. Astrophys.} {\bf 24},
127 (1986)

Press, W. H., Flannery, B. P., Teukolsky, S. A. \& Vetterling, W. T.
{\em Numerical Recipes} (Cambridge University Press, 1989)

Richardson, W. H. {\em J. Opt. Soc. America} {\bf 62}, 55 (1972)

Shannon, C. J. {\em Proc. I. R. E.} {\bf 37}, 10 (1949)

Skilling, J. \& Bryan, R. K. {\em Mon. Not. Roy. Astron. Soc.} {\bf 211},
111 (1984)

Snyder, D. L., Miller, M. I., Thomas, L. J., \& Politte, D. G. {\em IEEE
Transactions on Medical Imaging} {\bf MI-6}, 228 (1987)

\vspace{3ex}
{\Large {\bf Acknowledgements}\\
\normalsize

This work  has been supported  by contracts  ARC 94/99-178 ``Action de
Recherche Concert\'ee  de la Communaut\'e Fran\c{c}aise de Belgique'',
SC  005  ``Service   Center    and Research  Networks''     and P\^ole
d'Attraction Interuniversitaire P4/05 (SSTC, Belgium).\\[6ex]

\pagebreak

\begin{figure}[t]
\begin{center}
\leavevmode
\epsfxsize=16.0 cm 
%\epsffile{f1.eps}
\caption{Deconvolution of a  simulated image of  a star cluster partly
superimposed  on a  background   galaxy.   Top  left$\,$:  true  light
distribution  with  2  pixels  FWHM  resolution$\,$; bottom  left$\,$:
observed image  with 6  pixels  FWHM  and noise$\,$;   top middle$\,$:
Wiener  filter  deconvolution   of  the    observed image$\,$;  bottom
middle$\,$:  50    iterations  of   the   accelerated  Richardson-Lucy
algorithm$\,$;    top right$\,$:   maximum entropy  deconvolution$\,$;
bottom right$\,$: deconvolution with our new algorithm.}
\end{center}
\end{figure}

\begin{figure}[t]
\begin{center}
\leavevmode
\epsfxsize=16.0 cm
%\epsffile{f2.eps}
\caption{Deconvolution   of   a pre-discovery  image   of the
Cloverleaf gravitational    mirage  obtained  with the   ESO/MPI  2.2m
telescope  at La Silla (Chile).   Left$\,$: observed image with a FWHM
resolution of   1.3  arcsec$\,$;  right$\,$: our   deconvolution  with
improved sampling and a FWHM reso\-lution of 0.5 arcsec.}
\end{center}
\end{figure}

\begin{figure}[t]
\begin{center}
\leavevmode
\epsfxsize=16.0 cm
%\epsffile{f3.eps}
\caption{Deconvolution of an image  of the compact star cluster
Sk$\,$157  in the  Small  Magellanic Cloud.  Left$\,$: image  obtained
with the  ESO/MPI  2.2m telescope  at   La  Silla (1.1  arcsec  FWHM);
right$\,$: deconvolution with our algorithm (0.26 arcsec FWHM).}
\end{center}
\end{figure}

\begin{figure}[t]
\begin{center}
\leavevmode
\epsfxsize=16.0 cm
%\epsffile{f4.eps}
\caption{Simultaneous  deconvolution of simulated images.  Top
left$\,$: true light distribu\-tion with 2 pixels FWHM resolution$\,$;
top middle$\,$:   image   obtained with a    space  telescope$\,$; top
right$\,$:  image  obtained with  a  large ground-based telescope$\,$;
bottom  left$\,$:   sum of  the   two images$\,$;   bottom middle$\,$:
simultaneous  deconvolution    with   Lucy's     algorithm$\,$; bottom
right$\,$: simultaneous deconvolution with our new algorithm.}
\end{center}
\end{figure}

\begin{figure}[t]
\begin{center}
\leavevmode
\epsfxsize=16.0 cm
%\epsffile{f5.eps}
\caption{Simultaneous   deconvolution  of  4   simulated
adaptive-optics-like images.  Top  left$\,$: true  light  distribution
with  2 pixels FWHM resolution$\,$;   middle  and right$\,$: 4  images
obtained with    the   same instrument  but    in varying  atmospheric
conditions$\,$; bottom left$\,$:  simultaneous  deconvolution with our
new algorithm.}
\end{center}
\end{figure}

\begin{figure}[t]
\begin{center}
\leavevmode
\epsfxsize=16.0 cm
%\epsffile{f6.eps}
\caption{Intensity  ratio derived after deconvolution with the maximum
entropy method for a pair of point sources with variable separation.  
The true intensity ratio is 0.1, the peak S/N ratio is 100 and the 
original resolution is 7 pixels FWHM.}
\end{center}
\end{figure}

\begin{figure}[t]
\begin{center}
\leavevmode
\epsfxsize=16.0 cm
%\epsffile{f7.eps}
\caption{Photometric  test performed on a   synthetic  field
containing 200 stars with random positions and intensities, nearly all
blended  to  various degrees  (see text).    The  relative errors  are
plotted against the total intensity  (the latter being on an arbitrary
scale, corresponding  to an  integrated S/N  varying  from 10 to 400).
Open  symbols represent  heavily  blended stars  (the distance to  the
nearest neighbour is smaller than the FWHM), filled symbols correspond
to  less blended objects.     The dashed  curves are  the  theoretical
3$\sigma$ errors  for isolated stars,  taking into  account the photon
noise alone.}
\end{center}
\end{figure}

\begin{figure}[t]
\begin{center}
\leavevmode
\epsfxsize=16.0 cm
%\epsffile{f8.eps}
\caption{Astrometric test performed on the same  crowded field as in
Fig.~7.  The total  error  in position  (expressed in  fractions  of a
pixel)  is plotted  versus    the total  intensity.   Open   symbols
represent heavily  blended  stars  (the   distance to  the   nearest
neighbour is  smaller than the FWHM),  filled  symbols correspond to
less blended objects.}
\end{center}
\end{figure}

\begin{figure}[t]
\begin{center}
\leavevmode
\epsfxsize=16.0 cm
%\epsffile{f9.eps}
\caption{Deconvolution of an image containing 3 points sources
superimposed on a variable background.   Top left$\,$: observed image;
top  right$\,$: deconvolution with 50  iterations of the accelerated
Richardson-Lucy  algorithm; bottom  left$\,$:  deconvolution with 1000
iterations  of   the  two-channel  Lucy   method;    bottom right$\,$:
deconvolution with our algorithm.}
\end{center}
\end{figure}

\begin{figure}[t]
\begin{center}
\leavevmode
\epsfxsize=16.0 cm
%\epsffile{f10.eps}
\caption{Comparison  of  the background light distributions
deduced from  the deconvolution of  the  image  in Fig.~9,  using  the
two-channel Lucy method (top  left) and our  method (top  right).  The
bottom panels show the  square of the  difference between  the deduced
background (reconvolved   to   the same    2 pixels resolution    when
necessary) and the  exact solution, with  the two-channel  Lucy method
(left) and with our algorithm (right).}
\end{center}
\end{figure}

\end{document}